# Chapter 2: Fog Computing for Smart Grids: Challenges and Solutions

Linna Ruan[1,2], Shaoyong Guo[1], Xuesong Qiu[1], Rajkumar Buyya[2]
[1] Beijing University of Posts and Telecommunications, State Key Laboratory of Networking & Switching Technology, China
[2] Cloud Computing and Distributed Systems Laboratory, School of Computing and Information Systems, the University of Melbourne, Australia

**Abstract:**

Smart grids (SGs) enable integration of diverse power sources including renewable energy resources. They can contribute to the reduction of harmful gas emission, and support two-way information flow to enhance energy efficiency, along with small-scale Microgrids, acting as a promising solution to cope with environmental problems. However, with the emerging of mission-critical and delay-sensitive applications, traditional Cloud-based data processing mode becomes less satisfying. The use of Fog computing to empower the edge-side processing capability of Smart grid systems is considered as a potential solution to address the problem. In this chapter, we aim to offer a comprehensive analysis of application of Fog computing in Smart grids. We begin with an overview of Smart grids and Fog computing. Then, by surveying the existing research, we summarize the main Fog computing enabled Smart grid applications, key problems and the possible methods. We conclude the chapter by discussing the research challenges and future directions.

## 2.1 Introduction

In recent years, significant climate change, like global warming and air quality deterioration, threatens all the lives on the earth and attracts worldwide concern about harmful gas emission as well as energy issues. Given this background, the traditional power grids are transformed to Smart grids to enhance energy efficiency and system reliability, providing a promising solution to address environmental problems. Following this trend, Microgrids, as small-scale local power systems, also be proposed to optimize energy management individually or through collaboration with main grid. The two types of grids are implemented at the utility level and facility level respectively, while both contribute to energy system and environment. This chapter mainly focus on Smart grids, the large-scale conception. Due to many commonalities, most of the discussion also fits for Microgrids.

Smart grids enable two-way communication and integrate renewable resources for power generation, being used to support smart cities and other energy required scenarios. Apart from these benefits, it is subject to some problems during implementation, mainly reflected in four aspects. 1) A huge amount of data generated by Smart grid devices requires robust processing capability; 2) The emerging delay-sensitive applications propose instant response requirements; 3)Transmission of all data over the uplink increases the burden on the communication channels; 4) Uploading data to Cloud through the open Internet increases the risk of privacy violations.

The traditional mode of processing data in Cloud shows its limitations in this background, mainly due to the limited transmission resources and long response delay and, in particular, to the data privacy risk. Moving the processing of emergency data to the edge side is regarded as an efficient way to address these problems, which is also the main intention of Fog computing. Therefore, as one of the advanced technologies and the vertical downward extension of Cloud computing, Fog computing is discussed to be employed in Smart grids to enhance edge-side processing capability, reduce response time, relieve the burden of core network, and protect user privacy. During the application of Fog computing, two problems are deputed most. First, how to deal with the relationship between the two computing modes, should Fog computing replace or complement Cloud computing. Second, Fog computing benefits Smart grids on multiple aspects, while some new problems also emerged, how to cope with that.

In this chapter, we aim to conduct a comprehensive analysis of Fog computing in Smart grids. We begin with a brief introduction of Smart grids and Fog computing, focusing on their features, components, advantages and challenges. Then, we discuss the application of Fog computing in Smart grids by analysing the existing research and the current solutions, so as to illustrate the application scenarios and summarize the frequently used methods. Further, we outline the challenges of Fog enabled Smart grids and the future research directions. Finally, we conclude the chapter.

In general, we hope to clarify three problems through this chapter: Why is Fog computing suitable for Smart grids? What are the current solutions? And what challenges may exist for future applications?

**2.2 Smart Grids (SGs)**

SGs are defined in various ways by different organizations. In the United States, SGs are viewed as a large-scale solution to realize energy transformation from global network to the localized. While in China, SGs are defined as an approach that ensures energy supply based on physical network. For Europe, SGs mean a broader RE (renewable energy)-based system with society participation and countries integration [1]. Although there are differences in the definition of Smart grids, consensus has been reached on three aspects. 1) SGs are envisioned as the next generation electrical energy distribution network and an important part of smart cities. With the reliable communication system, SGs can manage energy more intelligently and effectively; 2) SGs allow two-way both electrical flow and information flow interaction between demand side and supply side, which makes energy consumption and pricing strategy easier to be monitored. In addition, supply-demand match, efficient energy utilization, and energy cost reduction can be realized; 3) SGs allow devices to interact information and is suitable for Internet of Things (IoT) scenarios. In a nutshell, Smart grids integrate advanced information and communication technologies into the physical power system to:

- enhance the level of system automation, hence contribute to operation efficiency;
- improve system security and reliability;
- fit the requirements of sustainable development better by using cleaner electricity resources and storage devices;
- enhance energy efficiency by facilitating two-way information interaction;
- allow customers to monitor their energy consumption and schedule electricity usage plan, which would benefit both themselves and the systems.

**2.2.1 Architecture**

A Smart grid can be described from three perspectives. First, from the perspective of functions, SGs can be viewed as the combination of physical infrastructures and information technologies.

Second, considering the core processes and participants, a Smart grid system consists of seven domains. Third, in view of the coverage scale, small-scale Microgrid as an important component of Smart grids has been hotly discussed in recent years. The details of these three ways are introduced as below.

**1) Perspective 1: Functions (Physical and information domains integration)**

The SG is a product of cross-domains integration, which distinguishes it from the traditional power grid. The physical domain refers to the electrical power systems, including the energy generation, transmission, distribution and consumption. While the information domain refers to information technologies, which are used to automatically transmit and retrieve data when necessary [2]. This integration is also reflected in interaction flows, relatively corresponding to electrical flows and information flows.

**2) Perspective 2: Core processes and participants**

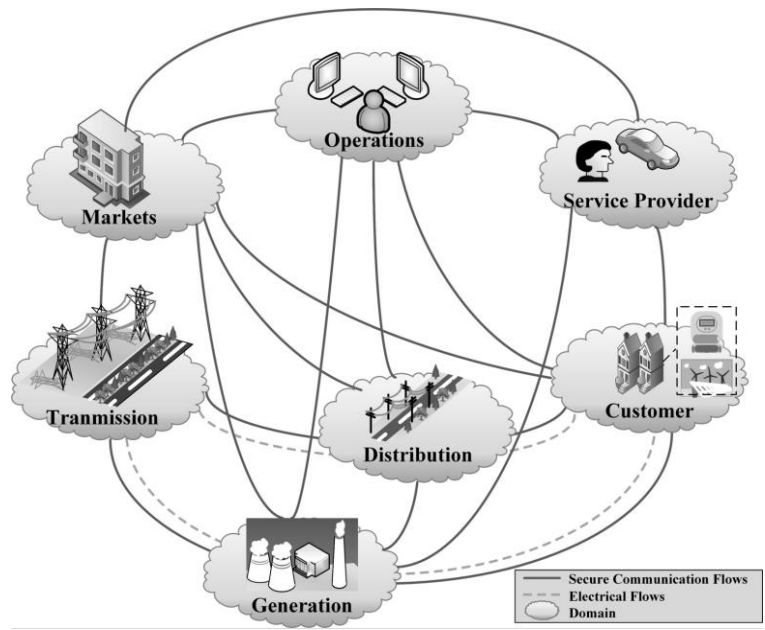

Figure 2.1: Conceptual model of Smart grids defined by NIST

Short ALT Text for Figure 2.1 [18 words]:

Seven domains, connected by secure communication flows and electrical flows, shows the major functions of a Smart Grid.

The National Institute of Standards and Technology (NIST) illustrates the functions of a SG with a conceptual model as shown in Figure 2.1, which defines seven important domains. The concept of each domain is explained as follows.

- Bulk generation domain

Generate electricity for later transmission, distribution and finally for residential, commercial or industrial use. The generation sources include traditional sources (such as fossil fuel, coal), and distributed energy sources (such as solar and wind power).

- Transmission domain

It is commonly defined as the carrier for long distance power transmission. In some specific scenarios, it also has the capability of electricity storage and generation.

- Distribution domain

Distribute electricity to or from (when the surplus power generated by distributed resources needs to be sent back to the market) customers. Similar with power transmission, the distribution domain has the capability of electricity storage and generation in some cases.

- Customers

They are the end users of electricity. According to the consumption habits and levels, they are divided into three types: residential, commercial, and industrial. Besides consuming electricity, customers may also generate and store electricity by embedding distributed resource infrastructures and batteries. In addition, demand-side management allows them to monitor and manage their energy usage.

- Service providers

Organizations that provide services for electricity users and utilities.

- Operations

The managers of the electricity movement.

- Markets

A trading place for operators and customers. In power grid systems, the markets are divided into whole-sale markets and retail markets, depending on the transaction mode.

**3) Perspective 3: Coverage scale**

By integrating distributed resources, a new form of SGs has been formulated, called as Microgrids. Extending the literature [3], its architecture is depicted in Figure 2.2.

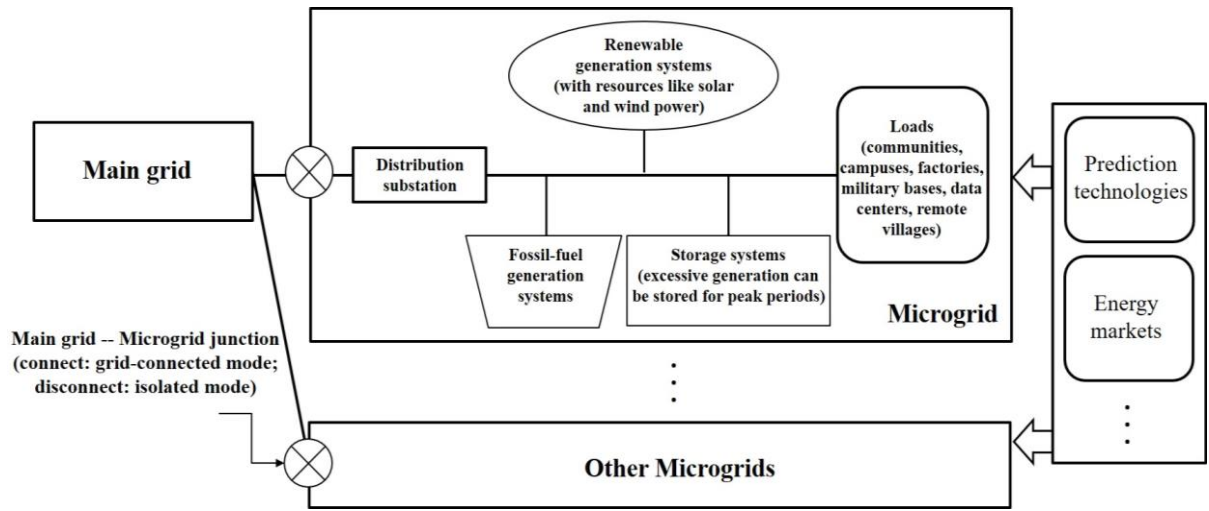

Figure 2.2: A Microgrid architecture

Short ALT Text for Figure 2.2 [36 words]:

Three parts, separated horizontally. The left two parts, organized with connect or disconnect mode, show the basic elements of Microgrids and their relationship with the Main Grid. The right part shows other elements associated with Microgrids.

Similar with SGs, Microgrids are defined by different organizations. For example, the Microgrid Exchange Group has defined a Microgrid as "a group of interconnected loads and distributed energy resources within clearly defined electrical boundaries that acts as a single controllable entity with respect to the grid". While the consensus on Microgrids lies in four aspects. 1) It is an important component of SGs [4]; 2) Compared with SGs, Microgrids refer to a smaller distributed local power system [4-5]; 3) Distributed generators and power-storage units are included; 4) It can operate either in conjunction with the main grid (excessive power that cannot be totally consumed locally can be sold to utilities through electricity market) or in an isolated mode (only provide services for end customers, which differentiates it from the centralized power generation form). Such obvious benefits listed below make Microgrids a hot topic for Smart grid researchers.

- Localized mode eases the distributed and renewable energy sources integration (such as solar and wind power), relieving the burden of generators in peak load periods and therefore reduce harmful gas emission.

- Locating close to demand side makes Microgrids easier to get the knowledge of uses' needs, resulting in efficiency increase and transmission cost reduction [6]. It provides better service by ensuring the energy supply of critical loads (loads generated by devices or organizations which require continuous power supply, such as military equipment, hospitals and data centers [2]), system quality, resilience and reliability, and allow customers to join demand-side management in an easier way.

- Provide strong support to the main grid by handling local urgent issues, such as the variability of renewables and the sensitive loads which require local generation, and providing auxiliary services to the bulk power system.

- Local-global controlling can be realized. With both regional requirements and overall performance taken into account, Microgrids offer better performance insurance and more opportunities for multi-technologies integration (electric vehicle, residential energy storage, rooftop photovoltaic systems and smart flexible appliances [2]), which is benefit for Smart grids development.

Many researchers have investigated the relationship between Smart grids and Microgrids. They are widely regarded as the implementation of new era grids at utility level and facility level respectively [4-5], [7]. Though with some differences on construction, both contribute significantly to energy system and environment.

**2.2.2 Current and Upcoming Problems**

**1) Latency requirements**

Emerging mission-critical and delay-sensitive SG applications, such as demand response, emergency restoration process and substation monitoring, require low round trip latency [8]. Based on characteristics, these applications can be categorized into two types, the flexible real-time ones, the fault tolerant but continuity-required ones. A typical application of the first type is demand-side management, which allows customers to monitor their electricity consumption in almost real time. However, Cloud computing cannot meet the requirements, due to long distance data transmission, possible channel congestions and server failures. The second type cannot be satisfied by Cloud computing either, because possible connection failures and processing delay can cause interruption, making it difficult for Cloud to support a continuity-required service.

**2) New security challenges**

While integrating various information technologies, SG systems are facing new security challenges. We categorize the challenges on device level, communication and system level.

- Device level—Security measures upgrade

*Resource-constrained devices.* SGs have many resource-constrained devices that hold limit capability to upgrade the security corresponding hardware and software in their lifespans. Moreover, a SG environment includes more participants, more technologies and frequently information interaction compared to traditional power grid, improving energy efficiency while being more vulnerable to security attacks. Hence, there is an urgent need to figure out an effective protection method for resource-constrained devices [9].

*Large number of devices.* As mentioned above, resource constraints pose a challenge to the security protection of devices. In some cases, connect to Cloud seems a proper way to upgrade the security credentials and software. But with the exponentially growing number of devices, it is impractical to allow all devices to do that, which is a resource, energy and time-consuming process. Therefore, in the face of large number of devices, how to ensure system security is still a big challenge.

- Communication and system level

*Cope with security problems while ensuring operation.* When faced with security issues, shut down-then-fix is the common way currently. However, shutdown also means interruption, which is intolerable to mission-critical or delay-sensitive services. Like the example proposed in [9], if an electric power generator chooses to shut down when met malware attack, severe disruption will be caused and lead to power outages. Therefore, fix-while-operate is a prospect mode and still a challenge for SGs.

*Keep private data be private.* In SGs, the collected data is usually stored in Cloud-based data centers for further processing or future use. Since the private information contained in the data set is valuable for making energy strategy and then making benefits, it is preferred by intruders, and even service providers or Cloud operators. Therefore, it is important to ensure transmission security during the way from smart meters to the central Cloud to prevent data leakage. However, the data is transmitted through open Internet and the number of connected users continues to increase, making it harder to figure out a strong privacy protection solution [10].

*Robustness.* Communication network is one of the adding components that differentiate SGs from traditional power grids. Hence, its robustness directly impacts how a Smart grid system would be judged. A robustness communication network means that it can keep normal or recover quickly even in such terrible situations, like natural disasters or human intervention. In this case, we care about whether there are advanced technologies that can be included to deal with emergencies intelligently in addition to existing resource-consuming solutions (providing redundant links or power backup facilities).

*Reliability.* Reliability has always been viewed as a challenge for one system especially for

SGs due to the high outages cost. According to a Sun Microsystems analysis, blackouts cost approximately US$1 million every minute of electric companies [8]. The main reasons lead to outages lie in three aspects. 1) Cannot get the accurate knowledge of system status in real-time; 2) Lack of prediction and analysis capabilities; 3) Lack of timely and effective response measures. SGs offer better communication, autonomous control, and management methods to relieve these problems. However, how to extend the current framework to handle diverse and sophisticated issues in the future still seems a problem.

**3) Distributed control**

As mentioned before, the basic components of SGs are geo-distributed, which is inefficient to be processed with remote centralized Cloud. Therefore, a distributed computing platform is preferred to provide location-based services and analytics, location-free billing and charging, and many more [11].

**4) Prediction responsiveness**

Prediction is the basis of making pre-decisions and its accuracy directly impacts whether a strategy proposed is appropriate. In a SG environment, demand and generation prediction are studied most. Demand prediction is divided into long-term prediction and short-term prediction according to the time interval. It is important for demand-side management. While generation prediction is usually used for renewable energy resources, such as solar panels and wind turbines. Its accuracy mainly depends on the weather prediction and it is an important component of Microgrids. In the past several years, Cloud platform is the carrier for the two types of predictions. However, instant decision-making is required for some specific applications recently and nearly real-time prediction is expected, which cannot be satisfied by Cloud computing. In this case, how to enhance prediction responsiveness while meeting the requirements of computing capability is a challenge.

**5) Supply-demand match**

Communication network enables information interaction between power providers and consumers. It enhances energy efficiency by supporting system balancing, and incentives customers to optimize their electricity usage by cutting or shifting peak periods demand. Supply-demand match intends to realize less energy waste and higher energy efficiency, benefit for both system and environment. However, the demand and renewable resource generation is always changing dynamically, which requires real-time information flow to support customer's immediate participation. The requirement cannot be satisfied by centralized Cloud processing due to large response delay. Given this background, how to implement real-time information interaction between providers and consumers is a problem.

**6) Complexity complicates the system management**

The SG is an increasingly complex system due to the quantity and rate sharply increased data and various technologies applied. Besides, there are some new services should be supported, such as two-way communications, real-time information interaction and demand-side management. Therefore, manage the complex system to realize all these functions, guarantee their requirements and balance the interests of all participants is really a challenge.

**2.3 Fog Computing-Driven Smart Grid Architecture**

Fog computing is first proposed by Cisco as the vertical downward extension of Cloud computing. By providing computing, communication, controlling and network storage capability at the proximity of data source, Fog computing contributes to response time reduction and complements edge-side processing capability. It is mainly used to handle mission-critical and delay-sensitive applications. From the tech giants to manufactures, Fog computing is discussed to be used in many scenarios, especially the Smart grids, which have such challenges mentioned above and view Fog computing as a proper technology to break the barriers. Before delving into the application of Fog computing in Smart girds, we give a brief introduction about the main features of Fog and discuss how to deal with the relationship between Fog computing and Cloud computing.

**2.3.1 Features**

The features of Fog can be simplified as AESR (Awareness, Efficiency, Scalability, Responsiveness) illustrated as follows. They also reflect the advantages of Fog computing on different aspects.

- Awareness

The awareness refers two aspects: objective awareness and location awareness. In a Smart gird, users' preferences are various, such as profit, Quality of Experience (QoE) and energy efficiency. Since Fog nodes are geographically distributed, each of which masters a relatively small area, making the nodes easier to get users' expectations, preferences, and provides suitable even customized strategies. It is like the general saying 'the right is the best', awareness drives Fog computing a good service provider.

- Efficiency

In a broader perspective, Fog computing is regarded as the added computing nodes between the end devices and the Cloud. Moreover, the capabilities of Fog computing are not limited to computing, communication and storage, these basic functions make it also a good resource manager and task scheduler. It integrates all the edge-side resources, such as smart appliances, computers, and find the best place for task processing with the combination of resource scheduling. In this way, Fog computing attains high efficiency in terms of both resource and system operation.

- Scalability

Fog platform locates close to users and small in size, making it easier to adjust according to environment requirements, supporting infrastructures update and scaling with less cost. In addition, Fog permits even small groups to access public programming interfaces, and copes with new emerging services well with good scalability.

- Responsiveness

Quick response is one of the main advantages of Fog computing and also the motivation for its proposal. Fog platforms implement data processing close to users, significantly shortening the transmission link, which makes actuators can obtain data analysis results and operation suggestions almost in real-time, meeting the requirements of mission-critical and delay-sensitive applications. This is essential for not only the Smart grid stable operation but also for the enable of millisecond reaction times of embedded AI to support emerging artificial applications, which is mentioned as one solution for applying Fog computing to Smart Grids in the next section.

### 2.3.2 Fog Computing Complements the Cloud

After introducing the features of Fog computing, we should explain why it still be proposed in the context of 'Cloud computing everything', and then clarify what kind of relationship exists between them. We will analyse from the following aspects.

- Latency

As mentioned before, geo-distributed Fog computing enables quick response due to locate proximity to end users. It offers users an opportunity to obtain analysis results timely and then go on operation or cope with urgent issues. As a comparison, centralized Cloud computing is a time-consuming process, caused by long distance data transmission both uplink and downlink. Moreover, huge amounts of data transmitted to Cloud puts great pressure on network channels, which may lead to congestion or even interruption. In a word, Fog computing can complement Cloud on real-time performance and reduce the possibility of channel congestions.

- Accessibility

Fog platforms locate at the edge side, enhancing the possibility for end devices to get served, especially for resource-constrained ones. Besides, Fog computing costs less either on time or energy compared to Cloud computing and processes data locally, protecting the system from channel congestions. In a conclusion, Fog complements Cloud on high accessibility, providing a more general and affordable solution for devices and services.

- Privacy

Data privacy protection, which means the protection of sharing confidential data with the third parties, is important for the reliability of a SG system. Similarly, we analyse the data processing mode of Fog and Cloud computing, so as to show their performance differences on data privacy. Fog computing enables data to be processed separately, indicating that private data can only be accessed by the Fog while public data also can be transmitted to the Cloud. While Cloud works on shared background [12] and data is transmitted through Internet, each link has a risk of data leakage, which endangers the safe and stable operation of SG systems. Therefore, Fog complements Cloud by providing another location for data, protecting data privacy while ensuring efficiency.

After the above analysis, it is easy to get a conclusion: Cloud has powerful computing capability, while Fog outperforms on latency, availability and privacy. Now we are able to answer the question "what's relationship exist between Cloud computing and Fog computing?". An appropriate answer is that Fog computing is a complement of Cloud computing. They have their own advantages, no one can replace the other. In a specific scenario, which paradigm to apply depends on the requirements of services and pursuing of users. Therefore, combining the centralized Cloud and the distributed Fog nodes to create a hybrid Fog-Cloud platform is the best way currently to address SG problems.

### 2.3.3 Fog Computing Helps Address Smart Grid Problems

After clarifying the relationship between Fog and Cloud, let's discuss one of the main application scenarios—Smart grids. The specific solutions will be introduced in the next section. While before that, we want to illustrate why Fog computing is suitable for Smart grids by comparing the earlier mentioned challenges faced by Smart grids (under traditional Cloud computing form) with the performance supplement that Fog can provide. The analysis can be summarized in Table 2.1.

Table 2.1: Fog provides effective ways to address Smart grid problems

| Classifications | Smart grid challenges | How Fog can help |
|---|---|---|
| Latency constraints | The flexible real-time applications and the fault tolerant but continuity-required applications have real-time response or continuous operation requirements, which cannot be satisfied by Cloud. | Fog locates at the proximity of end devices, strengthens edge computation, communication, controlling and storage capabilities, providing delay-reduced services, avoiding the risk of channel congestions, ensuring consistent operation. |
| New security challenges | On device level, continuous upgrading of security measures is hard to realize. On communication and system level, services continuity cannot be guaranteed. Private data faces the risk of leakage. Robustness and reliability are also not that satisfied. | Edge resource is empowered with Fog and able to support security infrastructures upgrade. Fog computing provides service with reduced delay which can ensure the continuous operation. Private data is processed at the edge, only public data can be further transmitted to the Cloud. |
| Distributed control | The basic components of SGs are geo-distributed. Centralized Cloud is highly cost and not that suitable. A distributed paradigm is preferred. | Fog computing follows distributed form, and is able to provide location-based services and analytics, location-free billing and charging. |
| Prediction | Demand and generation prediction are preferred in future SGs. Cloud can meet the requirement of computation while fail to update information in real-time. | Fog nodes have the computing capability to do basic prediction and can send back the results with short delay. It can catch the dynamic changes of information and update within latency limit. |
| Supply-demand match | Demand and renewable resource generation is always changing, while Cloud cannot process this frequently changing status. | Fog computing interacts the demand and pricing strategies between customers and providers timely, which facilitates demand response process. |
| Complexity | SG systems are becoming increasingly complex due to quantity and rate sharply increased data, various technologies applied and services should support. Complexity complicates the control of SGs. | Fog computing can undertake data analysing, support delay-sensitive services, relieve the burden of end devices, network and Cloud. Distributed mode means tasks can be split for processing, which decreases the complexity of SGs. |

## 2.4 Current Solutions for Applying Fog Computing to Smart Grids

A lot of research has discussed how to strengthen SGs with Fog computing. Since edge computing is interchangeably defined as Fog computing in most of the cases, both of the two computing paradigms in SG environments are discussed in this chapter and are uniformly called Fog computing. In this section, we depict a generic architecture for Fog enabled Smart grids (Fog-SGs), and review the mainly discussed services, key problems in strategy design and other technologies that may provide further performance improvement.

### 2.4.1 Fog-based Smart Grid Architecture

A Fog-based Smart grid architecture is proposed in Figure 2.3. It contains three layers, which are the infrastructure layer, constructed with residential, commercial and industrial buildings, acting as power demand side; the access layer, with Fog and Cloud computing, providing computing, communication and storage capabilities. Fog servers are deployed with base stations; and the supply layer, which is mainly responsible for power generation, transmission and distribution. The main services supported by Fog computing are also listed in the architecture and the details are shown as below.

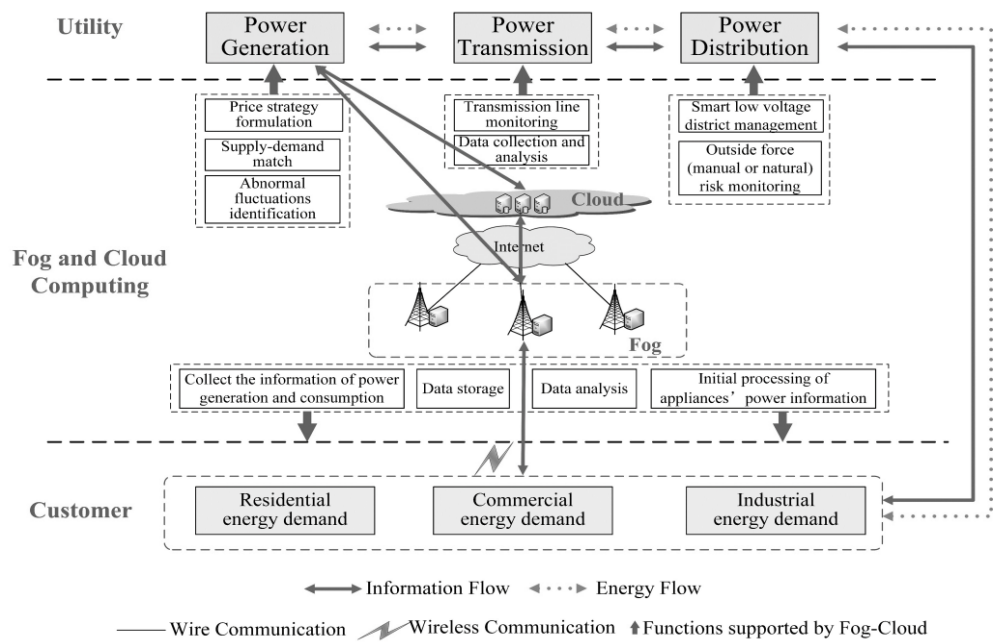

Figure 2.3: An architecture of Fog-Cloud enabled Smart grids

Short ALT Text for Figure 2.3 [34 words]:

Three layers, connected by information flow and energy flow, communicate via wire or wireless ways, show the major elements of Fog-Cloud enabled Smart grids and the functions can be supported by the new paradigm.

**2.4.2 Mainly Discussed Applications**

From the perspective of core links of SGs, we introduce how can Fog computing benefits the applications.

- **Power supply**

For power supply, Fog computing mainly be used to set price strategy, balance supply-demand and identify abnormal fluctuations. Demand-side management is one of the important applications in SGs, which has a potential of cutting off or shifting electric demand in peak periods. In this application, real-time pricing is the main motivation for customers, and supply-demand balancing is one of the aims. Due to the suitable computing and communication capabilities, and can provide delay-reduced services, Fog computing is envisioned as a well-suited technology to facilitate demand-side management. Besides, in [13], Fog nodes are considered to be embedded in charging points and detect the abnormal. It mentions that Fog computing can identify the abnormal status by analyzing sudden power fluctuations and then report to Cloud for further management.

- **Power transmission**

For power transmission, line status monitoring is a widely mentioned service that applies Fog computing [13-15]. Transmission line monitoring is important for getting full knowledge of the equipment condition (especially when met bad weathers, like high temperature, heavy rain, strong wind or snowstorm), supporting safe and stable operation of power systems. In this application, graph, video and data information is collected by unmanned aerial vehicles [13] or video sensors [15], which are controlled by edge network, then the information is sent to Fog nodes to filter and process. The analysis results reflect whether there is a possibility of failure, and effectively alleviates the bad effects as a result. In this process, quick response, data privacy and system reliability are really important, which can be better satisfied by Fog computing. Of course, if there is a large amount of data waiting for processing or the status is too complicated for Fog, hybrid Fog-Cloud will be adopted, in which pre-processing and further processing are considered to be executed respectively.

- **Power distribution**

For power distribution, Fog computing is considered to be used in smart low voltage (LV) district management and outside force (manual or natural) risk monitoring [16]. For LV district management, transformer terminal unit is enhanced with Fog computing which is used as low voltage side agent. LV topology identification, distribution fault diagnosis and line loss analysis therefore can be realized with less delay and has light pressure on storage and processing. For monitoring, the role of Fog is similar to when it is used for power transmission. In traditional mode, the collected data is stored in a local recorder and then sent to Cloud for processing. By facilitating Fog computing, lightweight data can be processed locally and warning information can be sent almost in real-time.

- **Substation**

For substation, the function of Fog computing is to monitor the operation status, equipment environment and analyze data, similar with power transmission process, while the difference is the source of data. In this scenario, the data mainly come from a large number of various sensors. With Fog computing, most of the information and data processing can be implemented locally and respond timely to ensure the warning before events, the suppression during events, and the reviews after events [13].

- **Power consumption**

For Microgrid systems, Fog computing is used to collect the information of power generation and users' electricity consumption in real-time, and then abstract their behavior mode based on power information in the time dimension. During system operation, the behavior mode is used to judge the power balance level and identify abnormal status [13].

For advanced metering systems, Fog devices are embedded in power meter concentrator to support the storage and analysis of data as well as services. The enhanced real-time interaction and price prediction capabilities empower the demand-side management application.

For smart home, most smart appliances require initial processing of power information. With time and cost-saving characteristics, Fog computing would be preferred to obtain better performance. In addition, Fog nodes can act as user agents, interacting information between users and Cloud, providing local data collection, operation status monitoring, small-scale controlling functions, contributing to emerging applications, such as demand response and fault diagnosis [16]. Table 2.2 summarizes the mentioned SG applications supported by Fog computing.

Table 2.2 SG applications supported by Fog computing

| Applications | Functions realized | Benefits provided | References |
|---|---|---|---|
| Power supply | Price strategy formulation (Demand prediction) | Support demand-side management. | [13] |
| | Supply-demand match | Enhance energy efficiency. | |
| | Abnormal fluctuations identification | Identify abnormal situations and send warning information timely; Enhance system reliability. | |
| Power transmission | Transmission line monitoring | Conceive the status of the equipment timely. | [13-15] |
| | Data collection and analysis | Ensure the warning before events, the suppression during events, and the reviews after events; Support safe and stable operation of power system. | |
| Power distribution | Smart low voltage (LV) district management | Fast topology identification, distribution fault diagnosis and line loss analysis. | [16] |

| | Outside force (manual or natural) risk monitoring | Lightweight data localized processing; Find abnormal situations and send warning information timely; Enhance system reliability. | |
| --- | --- | --- | --- |
| Power substation | Operation status and equipment environment monitoring | Get the condition of the operation status and equipment environment timely. | [13] |
| | Data collection and analysis | Ensure the warning before events, the suppression during events, and the reviews after events; Support safe and stable operation of power system. | |
| Power consumption | Power generation and users' electricity consumption information collection; Price prediction | Support demand-side management; Help users optimize electricity usage plan. | [13], [16] |
| | Abstract the behavior mode | Judge the power balance level and identify abnormal status. | |

**2.4.3 Key Problems Focused in Strategy Design**

Specific problems that are often discussed in Fog-SGs include: resource management, task scheduling, security and privacy protection, and comprehensive ones. The solutions provided are illustrated as follows.

- **Resource management**

A residential scenario is considered in [17]. It proposes a Cloud-Fog based SG architecture, in which multiple buildings are formulated as end user layer, and each building corresponds to one Fog device. By applying the Cloud-Fog mode, quick response can be realized. In addition, Particle Swarm Optimization (PSO), Round Robin and throttled algorithms are used to implement electric load balancing. It is verified that PSO outperforms the other two algorithms in terms of response time and total cost.

The work [18] focuses on a residential scenario and aims to optimize energy consumption, which is important for demand-side management application. A Cloud-Fog architecture is proposed and each Fog node manages energy demand scheduling of several buildings. It aims to reduce total energy cost and formulates the problem as a distributed cooperative demand scheduling game.

- **Task scheduling**

A Fog computing system is considered to provide strong storage and computing resources in SG communication network. The work [19] aims to minimize the total cost for the system running with subject to the tasks requirements. A green greedy algorithm is designed to provide a solution for the optimization problem.

Another work [20] takes SG communication network into consideration and designs a service caching and task offloading mechanism, to realize network load balancing and accelerate response. A computing migration model is also proposed to support task offloading from central Cloud to the edge network.

- **Security and privacy protection**

By facilitating information communication, such benefits like energy saving and customers' satisfaction improvement can be realized. However, it also means SG is more vulnerable to attacks which can be summarized as device attack, data attack, privacy attack and network attack [21]. Therefore, confidentiality, integrity, and availability (CIA), as the key judgments of a security policy should be highly guaranteed.

In [21], a Fog computing-based strategy is proposed to detect cybersecurity incidents in SGs. It models the anomaly detection process on the basis of generally used Open-Fog Reference Architecture. Sensors, actuators, smart meters and a central monitoring unit are included in the model and Fog is embedded in Phasor Measurement Units (PMUs) to do device measurement. The main process is: data acquisition, data preparation, features extraction (using k means algorithm) and behavior modeling, anomaly detection and score provision. Hence, unusual power usage can be detected, which is probable an attack on a SG system.

For failure recovery, a Fog computing-based dispatching model is proposed in [22]. In that paper, each regional power grid is in charge of a Fog node, which is responsible for fault information storage, repairing resource allocation and making dispatching plan with short delay. A dispatching algorithm based on genetic algorithm is carried out in Fog nodes, realizing cost reduction and satisfaction increase during repair process.

The work [23] presents a security Fog-SGs model, in which access authentication, data security and real time protection are described as expected functions. Then, for physical layer authentication, it adopts K-Nearest Neighbors (KNN) as a solution. A differential privacy data distortion technique consists of Laplace and Gaussian mechanisms is provided in [13].

- **Comprehensive problems**

Due to the complex environment of SGs, some research also tried to figure out solutions for comprehensive problems to balance multiple performance metrics, such as security, efficiency and functionality. [24] builds a Fog computing-based SG model and proposes a concrete solution for both aggregation communication and data availability. By encrypted under a double trapdoor cryptosystem, security data aggregation is implemented. The solution is designed for service providers to realize dynamic control and electricity distribution. Similarly, a privacy-preserving authentication and data aggregation scheme for Fog-based Smart grids is proposed in [25]. In that scheme, short randomizable signature and blind signature are used for anonymous authentication, and then Fog nodes are used to solve billing problems. Table 2.3 summarizes the mentioned problems and solutions.

Table 2.3 Key problems and solutions

| Classification | Problems | Objectives | Solutions | References |
|---|---|---|---|---|
| Resource management | Electric load balancing | Reduce response time and total cost | Build a Cloud-Fog based SG architecture; Propose an algorithm based on Particle Swarm Optimization (PSO). | [17] |
| | Energy demand scheduling | Reduce total energy cost | Build a Cloud-Fog based SG architecture; Formulate a distributed cooperative demand scheduling game. | [18] |
| Task scheduling | Task scheduling | Minimize the total cost for system running | Propose a green greedy algorithm. | [19] |
| | Service caching and task offloading | Balance network load and reduce communication delay | Propose a computing migration model; A load-balancing algorithm based on popularity and centrality. | [20] |
| Security and privacy protection | Cybersecurity incidents detection | Detect anomalies and reduce communication delay | Propose an anomaly detection process model and a detection method based on Fog computing. | [21] |
| | Failure recovery | Reduce cost and increase satisfaction during repair process | Propose a dispatching model based on Fog computing and a dispatching algorithm based on genetic algorithm. | [22] |
| | Access authentication, data security, real time protection | Realize physical security and improve energy efficiency | Propose a secure Fog-SGs model, a physical security approach and an electricity forecasting method. | [23] |
| Comprehensive problems | Aggregation communication and data availability | Realize dynamic control and electricity distribution | Propose a Fog computing-based SG model; A concrete solution for both aggregation communication and data availability. | [24] |
| | Security and privacy issues in Fog-based Smart grids communication | Anonymous authentication and billing | Propose a privacy-preserving authentication and data aggregation scheme for Fog-based Smart grids. | [25] |

**2.4.4 Fog+**

Fog complements Cloud in multiple aspects, while for a Fog-SGs system, it is not enough. By integrating other advanced technologies, the performances are possible to be further improved. We select three hot discussed technologies, which are blockchain, AI and SDN. They are expected to empower energy security and privacy protection, prediction and resource flexibility for Fog-SGs respectively.

- **Fog + Blockchain**

As mentioned before, Fog computing assists SGs security on transmission, service continuous operation. Nonetheless, the security and privacy protection of Fog-SG systems is still a challenge, due to Fog server itself is less secure than Cloud and there exist interaction and service migration between heterogeneous Fog nodes [26]. Hence, this problem has been stressed in recent years, and some security solutions are proposed.

However, since SGs have many resource-constrained devices, most of the conventional methods (certificate authority-based, ring signature, blind signature, group signature) are not suitable, due to bad traceability and participation flexibility, high computation and communication overhead. Given this background, blockchain is envisioned as a new chance for SGs. Blockchain allows network participants to record system in a distributed shared ledger. In a blockchain enabled system, Fog servers need to join while end uses are not required to, which prevent users' identity leakage. At the same time, the change in the participation status of one user will not influence others, as the registration is identity-based. The smart contract included also supports traceability and revocability [27]. Therefore, blockchain is chosen as fitting all the security requirements of SGs, and its scalability requirements can be complemented by Fog computing. Moreover, both blockchain and Fog computing follow decentralized network form, which makes them possible to integrate.

In [27], a blockchain-based mutual authentication and key agreement protocol is proposed for Fog-SG systems. In addition to providing basic security properties, it also offers an efficient method for key update and revocation as well as conditional identity anonymity with less costs.

From another perspective, [28] proposed a permissioned blockchain edge model for Smart grid network. It focuses on privacy protections and energy security issues, and formulates an optimal security-aware strategy by smart contracts.

- **Fog + AI**

Smart grids require accurate prediction of demand, pricing and generation capability to support resource preparing or policy setting and contribute to SG applications, such as demand-side response and monitoring. The basic elements of prediction are huge amounts of data, computation capability and intelligent algorithms. Evaluation metrics are speed, accuracy, memory and energy [29]. The amount of data generated by the SG infrastructures increases sharply both in terms of quantity and rate, which could be the first element of prediction. Fog enhances edge side computation capability and can provide delay-reduced services, match the second element. Then, for the third element, artificial intelligence algorithms seem to be the best choice currently. The following two cases illustrate how AI is supported by fog computing and how it is applied in SGs.

The work [16] focuses on distribution outside force damage monitoring application. For the processing platform, the fast and accurate vehicles identification with AI is the most critical segment. Since the AI algorithm is a resource-consuming process, the hybrid Fog-Cloud is envisioned as the most suitable technology to enhance its operation performance and save system resources.

Demand and dynamic pricing predictions are considered in [13]. Demand prediction is important as letting providers understand the needs of users, guiding the generation plan and contribute to electric resource balancing. Dynamic pricing acts as the main incentive for customers to respond to demand shifting or cutting off in peak period, and therefore directly relates to users' benefits. The suitable AI approaches are listed as Auto Regressive Integrated Moving Average models, Auto Regressive models, ANN, Fuzzy logic and (Long short-term memory) LSTM in that paper.

- **Fog + SDN**

Fog computing offers SGs benefits on multiple aspects. However, for the increasing scale of SG networks, how to transmit data to the Fog servers or sometimes to the Cloud could bring significant effect on resource efficiency. To face the challenge of data routing, involvement of Software Defined Networking (SDN) is viewed as a potential solution of Fog-SGs, mainly due to driving a more manageable and flexible network with a global view brought by decoupling the network control plane and data plane. [30] proposed an SDN-based data forwarding scheme for Fog-enabled Smart grids. With the global information provided by SDN, the shortest path is calculated and a path recovery mechanism is designed to avoid link failures. SDN is mainly used to facilitate multicasting and routing schemes for SGs [30].

## 2.5 Research Challenges and Future Directions

### 2.5.1 Security and Privacy

As mentioned before, the security of Fog computing is still a challenge, also for Fog-SGs. The reasons are threefold: 1) Heterogeneous. As the emerging of Internet and SGs, Fog computing will be combined with various technologies and access to other systems. Interact information and migrate services between heterogeneous devices in large or small scales face the danger of malicious attacks; 2) Dynamism and openness. Fog computing is a small edge-side computing platform and runs in an open and dynamic environment, which is vulnerable to attacks; 3) Data dispersion. Since each Fog node has limited capability, data usually be transmitted to geo-distributed nodes to store or process. It brings a risk of data leakage, packets loss or incorrect organization, breaking data integrity.

In conclusion, security issues of Fog-SGs, such as edge control, data dispose, computation and communication still need new ideas to match the distributed, heterogeneous and complex environment. Blockchain is a probable method to address identity risks, while the research is still at the initial stage. Scalable enhancement, layered mechanisms are the research directions and how to deal with out-sourcing services as well as off-chain status need further discussion.

## 2.5.2 Huge Amounts of Data Processing

With the emerging of Smart grid infrastructures, various system data increases both in rate and amount, which brings high complexity and aggravates the burden for data processing. The challenges brought by huge amounts of data are mainly reflected in the following aspects.

- **Data control, security and privacy**

The quantity, diversity and rate of data in SGs increases dramatically. Based on broad consensus, Fog computing strengthens the computation, communication, controlling and storage capability of edge side and is envisioned as a good choice to cope with the data explosion of SGs. However, a more accurate statement is that Fog can relieve the problems, not solve. Due to the characteristics, Fog computing platform means a set of capacity limited nodes. Indeed, the nodes can interact to support powerful data processing, but either for tech giants or manufacturers, it is really challenging. We will analyze the reasons from different stages of data processing. First, data split. When the resource of one Fog node is not sufficient to hold the data collected by sensors, advanced metering infrastructure (AMI) or other devices, data will be split into several parts to process. However, what split granularity should be selected. Large granularity is benefit for organization and data integrity, while small granularity can use resource with high efficiency and is more likely to be processed quickly. In this case, the choice seems depend on what performance we pursue. Is it easy? Don't get the conclusion too early, let's continue. Then, data distribution. This stage contains three problems, which node to send, how much to send and which path to select. For the first problem, the node with the richest available resources, or closest distance, or with the best comprehensive performance in combination with the degree of credibility always be selected as the destination. For the second problem, it is usually decided along with the first one, depends on what we pursuing for, the system balance, the short delay, the energy efficiency or other else. Now for the third problem, two choices usually be mentioned, the shortest one or the most security one. Finally, data storage. Where to store and how much to store. Should the data be sent back to end device, retained at the Fog node or sent to the Cloud. We also have to make a decision.

Now, it is easier to reach a conclusion. During data processing, there are several problems should be considered in each stage, and each problem has several choices. It seems that these choices all depend on our goals, but the goals of different stages may not be the same. To achieve the best comprehensive performance, we must make various considerations, trades off between different performances with subject to the requirements of specific services. In the existing research, the solutions proposed are usually based on a specific application with such assumptions, which has limited value for a real industrial scenario. Therefore, for the data processing of Fog-SGs, how to deal with more and more complicated situations is very challenging and need more discussion.

- **Artificial intelligence integration**

Prediction is one of the important aspects to show the smart of SGs. As we discussed, large amounts of data collected from SG devices is the basis of prediction, and the artificial intelligence (AI) algorithm is an essential tool to guarantee the accuracy. However, since the AI algorithms are resource-consuming, in Fog enabled Smart grid environments, some aspects are still challenging. 1) Limit capability. Considering the complexity of prediction, it is usually carried out in Cloud traditionally. However, due to the emerging of applications and higher requirements proposed, real-time prediction is expected in Smart girds. Fog computing can realize quick response for most of the services, but it is not sure for prediction, depending on the complexity of the algorithm. Therefore, the edge side still needs other technologies to facilitate capability of Fog-SGs; 2) Follow the dynamics of algorithms. AI algorithms are not that stable, which means that they are more like an art, designed for the specific problem and innovation always come out, including more parameters, more layers, or new structures. Enable Fog-SG systems keeping up with new AI designs is obvious another challenge.

### 2.5.3 Fog and Cloud Combination

The hybrid Fog-Cloud computing is envisioned as the most promising mode to be integrated in SGs. But how to collaborate the two kinds of paradigms to show their own best and also realize the complement is really a challenge. Resource management problems are discussed most, which contains resource aggregation, task offloading, caching and storage. Due to virtualization technologies, the resource can be customized as containers or virtual machines, which enhances the resource flexibility but makes it more complex to formulate resource management strategies. Besides, the participation process of data and applications complicates the problem further. In the future, how to make generic strategies and customized strategies for specific SG scenarios with given optimization objectives still need more discussion.

### 2.5.4 Fog Devices Deployment

Fog computing has been discussed about how to be integrated into Smart grids in recent years. However, most of the solutions have such assumptions and not suitable for the real status. Consider about the deployment of Fog computing, there are usually two choices. 1) Embedding Fog computing functions in the existing Smart grid equipment. 2) Design new Fog computing equipment. The first option costs less, but requires manufacturers' permission for the equipment transforming. It is necessary to ensure the performance of the embedded part, especially the safety performance, while not bringing negative impact on the existing performances. The second option costs higher, but it does not need to negotiate with the device manufacturers, just needs to consider the access with other devices. But, both of the options require the support of the government and power suppliers, which control devices accessing and ensure the stable operation of Smart grids. As discussed above, the deployment and realization of Fog functions in a real SG is still a challenge.

## 2.6 Summary and Conclusions

In this chapter, we investigated the application of Fog computing in Smart grids from three perspectives: characteristics, solutions and challenges. Through the brief introduction to the components and features of Smart grids and Fog computing, two problems are analyzed. (1) Why Fog computing still be proposed in the context of 'Cloud computing everything' and how to deal with the relationship between Fog and Cloud computing? A proper answer is: Fog computing is proposed as a complement of Cloud computing. They have their own advantages, no one can replace the other. Under a specific scenario, which paradigm to apply depends on the constraints of devices, requirements of services and users. The hybrid Fog-Cloud computing is a generic and the most promising mode to be integrated in future SGs. (2) Why Fog computing is suitable for Smart grid systems? The answer is: the challenges faced by Smart grids (under traditional Cloud computing form), such as latency constraints, new security issues, distributed control, prediction, demand-supply match and complexity problems, all can be relieved with the performance supplement that Fog provides.

Through the analysis of the existing research, we identified the state-of-the-art of Fog-SGs. A generic architecture is proposed and the main applications of Fog computing are discussed from the perspective of core links of Smart grids. In summary, Fog computing is mainly used for delay-sensitive or mission-critical applications, to support prediction, monitoring and information interaction. These functions are the basis of Microgrids, demand-side management and communication network, which are the emerging areas in Smart grids.

From the perspective of theoretical research, hot issues in policy design are also introduced. They are divided into resource-related, task-related, security-related and comprehensive issues, which usually be summed up as an optimization problem with such constraints. The optimization objectives mainly include delay, energy consumption, satisfaction, system balancing, resource efficiency, safety and credibility, etc. Since applications in Smart grids usually involve multiple participants and their benefits need to be balanced, game theories are commonly used to solve these problems. In addition, optimization algorithms such as genetic evolution algorithms are widely used as solutions.

In response to the evolvement of industrial applications and technologies, we also proposed research challenges and future directions for Fog enabled Smart grids from the aspects of security and privacy, huge amounts of data processing, Fog and Cloud combination, and Fog devices deployment.

In summary, this chapter has analyzed the advantages and potential value of applying Fog computing in Smart grids. Considering many commonalities, many concepts presented in this chapter also apply to Microgrids. Obviously, Fog computing acts as a strong tool for designing optimized Smart grids systems that can fulfill the emerging requirements of applications with its outstanding "compute, communicate, storage, control" capability.

**Figure Legend**

Figure 2.1: Conceptual model of Smart grids defined by NIST

Figure 2.2: A Microgrid architecture

Figure 2.3: An architecture of Fog-Cloud enabled Smart grids

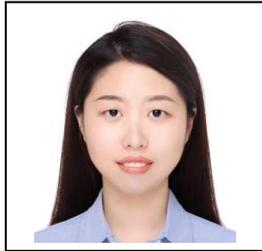
Linna Ruan received the B.S. degree in communication engineering from the Beijing Information Science and Technology University, Beijing, China, in 2016. She is working toward the Ph.D. degree in information and communication engineering with the Beijing University of Posts and Telecommunications, Beijing, China. As a visiting Ph.D. student, she is currently studying in the Cloud Computing and Distributed Systems (CLOUDS) Laboratory at the University of Melbourne, Australia. Her research interests lie in edge and cloud computing for power and Internet of Things (IoT) scenarios, including resource allocation, service computation offloading, etc.

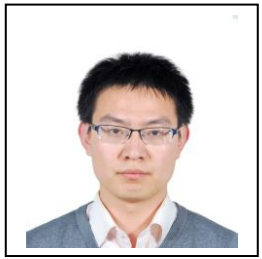
Shaoyong Guo received the Ph.D. degree from the Beijing University of Posts and Telecommunications, Beijing, China. He is currently with the Department of State Key Laboratory of Networking and Switching Technology, Beijing, China. His research interests include blockchain application technology, edge computing, and IIoT in energy Internet. He takes the lead to declare three ITU-T standards and wins several Provincial and Ministerial Prizes.

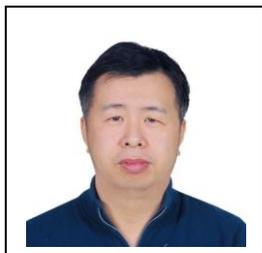
Xuesong Qiu (Senior Member, IEEE) was born in 1973. He received the Ph.D. degree from the Beijing University of Posts and Telecommunications, Beijing, China, in 2000. He is currently a Professor and the Ph.D. Supervisor with the State Key Laboratory of Networking and Switching Technology, Beijing University of Posts and Telecommunications. He has authored about 100 SCI/EI index papers. He presides over a series of key research projects on network and service management, including the projects supported by the National Natural Science Foundation and the National High Tech Research and Development Program of China.

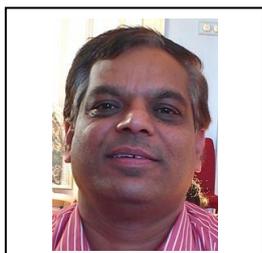
Dr. Rajkumar Buyya is a Redmond Barry Distinguished Professor and Director of the Cloud Computing and Distributed Systems (CLOUDS) Laboratory at the University of Melbourne, Australia. He is also serving as the founding CEO of Manjrasoft, a spin-off company of the University, commercializing its innovations in Cloud Computing. He has authored over 750 publications and seven text books including "Mastering Cloud Computing" published by McGraw Hill, China Machine Press, and Morgan Kaufmann for Indian, Chinese and international markets respectively. Dr. Buyya is one of the highly cited authors in computer science and software engineering worldwide (h-index=137, g-index=304, 99, 800+ citations).

Dr. Buyya is recognised as Web of Science "Highly Cited Researcher" for four consecutive years since 2016, IEEE Fellow, Scopus Researcher of the Year 2017 with Excellence in Innovative Research Award by Elsevier, and the "Best of the World", in Computing Systems field, by The Australian 2019 Research Review. Software technologies for Grid, Cloud, and Fog computing developed under Dr.Buyya's leadership have gained rapid acceptance and are in use at several academic institutions and commercial enterprises in 50 countries around the world. He served as founding Editor-in-Chief of the IEEE Transactions on Cloud Computing. He is currently serving as Editor-in-Chief of Software: Practice and Experience, a long standing journal in the field established ~50 years ago. For further information on Dr.Buyya, please visit his cyberhome: www.buyya.com.